# Unitary Transformation of Two-Dimensional Spin-Orbit Coupled Models

Manish Kumar Mohanta

Department of Physics, School of Basic Sciences, Indian Institute of Technology Bhubaneswar, Bhubaneswar 752050, Odisha, India

Email: manishkmr484@gmail.com, mkmohanta@iitbbs.ac.in

**Abstract:** The Rashba, Dresselhaus, and Weyl Hamiltonians form a foundational framework for modeling spin-orbit interactions across condensed matter systems. Although they describe distinct material classes and produce seemingly different spin textures, they are conventionally treated as separate, unrelated theoretical frameworks. Here, this work demonstrates that the linear 2D Rashba and Weyl models are connected by a specific unitary transformation that maps one Hamiltonian exactly onto the other. The same unitary can be applied to map the linear Dresselhaus-1 model onto the Dresselhaus-2 models and vice versa. Such hidden correspondence establishes a unified theoretical foundation for spin-orbit interactions, deepening our conceptual understanding of spin-orbit coupling and opening new avenues for exploring complex spin textures. To illustrate the application, this work introduces a unique, improved, and more realistic model Hamiltonian $\mathcal{H}_{MKM}$ combining all known foundational spintronic models, where the stringent condition of equal spin-orbit coupling strength of Rashba and Dresselhaus may not be required to observe persistent spin texture under MKM transformation.

**1. Introduction:** The exploration of spin-orbit coupling (SOC) has been a cornerstone of modern condensed matter physics, leading to the discovery of profound phenomena such as topological insulators, the spin Hall effect, and persistent spin textures. Among the various manifestations of SOC, the Rashba and Weyl Hamiltonians represent two paradigmatic models that give rise to distinctive electronic band structures with profound consequences for materials properties. [1–3] While both models describe momentum-dependent spin splitting, their fundamental origins, mathematical structures, and the resulting physical phenomena are markedly different.

The Rashba effect arises from structure inversion asymmetry (SIA), typically at surfaces or interfaces where the inversion symmetry is inherently broken. The canonical form of the 2D Rashba Hamiltonian is:

$$H_R(k) = \frac{\hbar^2 k^2}{2m^*} + \alpha(\sigma_x k_y - \sigma_y k_x) \dots\dots\dots\dots\dots (1)$$

Here, the first term represents the standard parabolic kinetic energy with the effective mass $m^*$ and the second term is the Rashba SOC, characterized by the coupling strength $\alpha$. The Pauli matrices $\sigma_x$ and $\sigma_y$ ensure that the spin of an electron is locked in a direction perpendicular to its momentum, $k = (k_x, k_y)$. The eigenvalue of this Hamiltonian describes two spin-split parabolic bands. Rashba systems exhibit helical spin textures on parabolic Fermi surfaces, with splitting tunable by gate voltage, yet they are topologically trivial unless additional terms are included. This model accurately describes semiconductor quantum wells (InGaAs/InAlAs), oxide interfaces (LaAlO$_3$/SrTiO$_3$), surfaces of topological insulators (Bi$_2$Se$_3$), and a growing family of 2D monolayers, as well as van der Waals heterostructures. [4–10]

In contrast, the Weyl Hamiltonian provides one of the most fundamental low-energy descriptions of relativistic quasiparticles in condensed-matter systems. While three-dimensional Weyl semimetals host isolated Weyl nodes protected by topology, reduced-dimensional counterparts have attracted increasing attention due to their relevance in two-dimensional (2D) quantum materials and engineered heterostructures. In 2D, a Weyl-type model describes linearly dispersing band crossings governed by spin-momentum locking, serving as an effective framework to capture chiral fermions, unconventional spin textures, and symmetry-protected degeneracies. The simplest form of a Weyl Hamiltonian can be written as;

$$H_W(k) = \hbar v_F(\sigma_x k_x + \sigma_y k_y) \dots\dots\dots\dots (2)$$

where $v_F$ is the Fermi velocity. This Hamiltonian hosts a Dirac-like linear dispersion but lacks inversion or time-reversal symmetry. The resulting eigenstates exhibit a helical spin texture, with spin orientation locked perpendicular to momentum. The key physical consequence of this model is the emergence of topological transport phenomena, most notably the anomalous Hall effect, which arises from the Berry curvature associated with the Weyl points. [11] Protected linear band crossing Weyl nodes emerge in systems such as monolayer bismuthine on SiC [12], α-antimonene [13], and certain Kagome lattices [14–16].

The distinguishing features of these two models translate directly into spintronic functionality. The Rashba interaction enables the Datta-Das spin field-effect transistor [17,18], electrical control of spin precession, and efficient spin-to-charge interconversion via the inverse spin-galvanic effect. Weyl nodes with their giant Berry curvature promise dissipationless edge currents and large anomalous Hall angles at the nanoscale- key ingredients for ultra-low-power memory and logic devices. Focusing on the linear 2D models, this work elucidates previously unrecognized correspondences between different SOC models.

## 2. Unitary Operator and Transformation of Linear 2D Hamiltonians:

**2.1 Correspondence between Rashba and Weyl Models:** A unitary operator $U(\phi) = e^{-i\frac{\phi}{2}\sigma_z}$ that implements the rotation in spin space about the $z$-axis maps the linear 2D Rashba Hamiltonian onto the Weyl Hamiltonian at $\phi = \frac{\pi}{2}$.

**Proof-1** The 2D Rashba Hamiltonian in the first order in reciprocal space is represented by:

$$\mathcal{H}_R(k) = (\sigma_x k_y - \sigma_y k_x) \quad \ldots\ldots\ldots\ldots (3)$$

The unitary operator that implements the rotation in spin space about the $z$-axis by an angle $\phi$ is given by;

$$U(\phi) = e^{-i\frac{\phi}{2}\sigma_z} \quad \ldots\ldots\ldots\ldots (4)$$

Under the spin rotation angle $\phi$, Pauli matrices transform as;

$$U(\phi)\sigma_x U(\phi)^\dagger = \cos\phi\ \sigma_x + \sin\phi\ \sigma_y \quad \ldots\ldots\ldots\ldots (5)$$

$$U(\phi)\sigma_y U(\phi)^\dagger = -\sin\phi\ \sigma_x + \cos\phi\ \sigma_y \quad \ldots\ldots\ldots\ldots (6)$$

Now applying a rotation of $\phi = \frac{\pi}{2}$, the Pauli spin matrices transform as follows:

$$U\left(\frac{\pi}{2}\right)\sigma_x U\left(\frac{\pi}{2}\right)^\dagger = \sigma_y \quad \ldots\ldots\ldots\ldots\ldots (7)$$

$$U\left(\frac{\pi}{2}\right)\sigma_y U\left(\frac{\pi}{2}\right)^\dagger = -\sigma_x \quad \ldots\ldots\ldots\ldots\ldots (8)$$

Now, applying the unitary operator to the 2D Rashba Hamiltonian, the new transformed Hamiltonian is given by;

$$\mathcal{H}_T(k) = U(\phi)\mathcal{H}_R(k)U(\phi)^\dagger \quad \ldots\ldots\ldots\ldots\ldots (9)$$

For $\phi = \frac{\pi}{2}$ rotation, the transformed Hamiltonian $\mathcal{H}_T$ can be rewritten as:

$$\mathcal{H}_T(k) = U\left(\frac{\pi}{2}\right)\mathcal{H}_R(k)U\left(\frac{\pi}{2}\right)^\dagger = (U\sigma_x U^\dagger)k_y - (U\sigma_y U^\dagger)k_x$$

$$\mathcal{H}_T(k) = (\sigma_y k_y + \sigma_x k_x) = \mathcal{H}_W$$

The right-hand side is precisely the linear 2D Weyl SOC term.

**Proof-2** Matrix Method:

The Matrix forms of Rashba and Weyl Hamiltonians are given by:

$$\mathcal{H}_R = \begin{pmatrix} 0 & k_y + ik_x \\ k_y - ik_x & 0 \end{pmatrix}$$

$$\mathcal{H}_W = \begin{pmatrix} 0 & k_x - ik_y \\ k_x + ik_y & 0 \end{pmatrix}$$

The spin rotation by $\phi = \frac{\pi}{2}$ about the z-axis is

$$U\left(\frac{\pi}{2}\right) = e^{-i\frac{\pi}{4}\sigma_z} = \begin{pmatrix} e^{-i\pi/4} & 0 \\ 0 & e^{+i\pi/4} \end{pmatrix}$$

Its Hermitian conjugate is

$$U\left(\frac{\pi}{2}\right)^\dagger = \begin{pmatrix} e^{+i\pi/4} & 0 \\ 0 & e^{-i\pi/4} \end{pmatrix}$$

Now let's compute $\mathcal{H}_T(k) = U\mathcal{H}_R U^\dagger$

$$U\mathcal{H}_R = \begin{pmatrix} e^{-i\pi/4}.0 & e^{-\frac{i\pi}{4}}(k_y + ik_x) \\ e^{+\frac{i\pi}{4}}.(k_y - ik_x) & e^{+i\pi/4}.0 \end{pmatrix} = \begin{pmatrix} 0 & e^{-\frac{i\pi}{4}}(k_y + ik_x) \\ e^{+\frac{i\pi}{4}}.(k_y - ik_x) & 0 \end{pmatrix}$$

Multiplying $U^\dagger$ on the right side:

$$\mathcal{H}_T(k) = (U\mathcal{H}_R)U^\dagger = \begin{pmatrix} 0.e^{+i\pi/4} & e^{\frac{-i\pi}{4}}(k_y + ik_x).e^{-i\pi/4} \\ e^{+\frac{i\pi}{4}}.(k_y - ik_x).e^{+i\pi/4} & 0.e^{-i\pi/4} \end{pmatrix}$$

Combining the phase factors: $e^{-i\pi/4}.e^{-i\pi/4} = e^{-i\pi/2} = -i$ and $e^{+i\pi/4}.e^{+i\pi/4} = e^{+i\pi/2} = +i$.

$$\mathcal{H}_T(k) = \begin{pmatrix} 0 & -i(k_y + ik_x) \\ +i(k_y - ik_x) & 0 \end{pmatrix} = \begin{pmatrix} 0 & k_x - ik_y \\ k_x + ik_y & 0 \end{pmatrix} = (\sigma_x k_x + \sigma_y k_y) = \mathcal{H}_W$$

Through two complementary derivations, this work shows that the 2D Rashba and Weyl SOC terms are connected by a $\frac{\pi}{2}$ rotation about the z-axis. The spin texture obtained from the Hamiltonian $\mathcal{H}_T(k) = U(\phi)\mathcal{H}_R(k)U(\phi)^\dagger$ under a different rotation angle $\phi$ is shown in Figure 1. Figure 1 also supports the derivation above that at a rotation angle, $\phi = \frac{\pi}{2}$, the Rashba spin texture is transformed into Weyl spin textures. And for any other angle, $\mathcal{H}_T(k)$ generates a spiral spin texture.

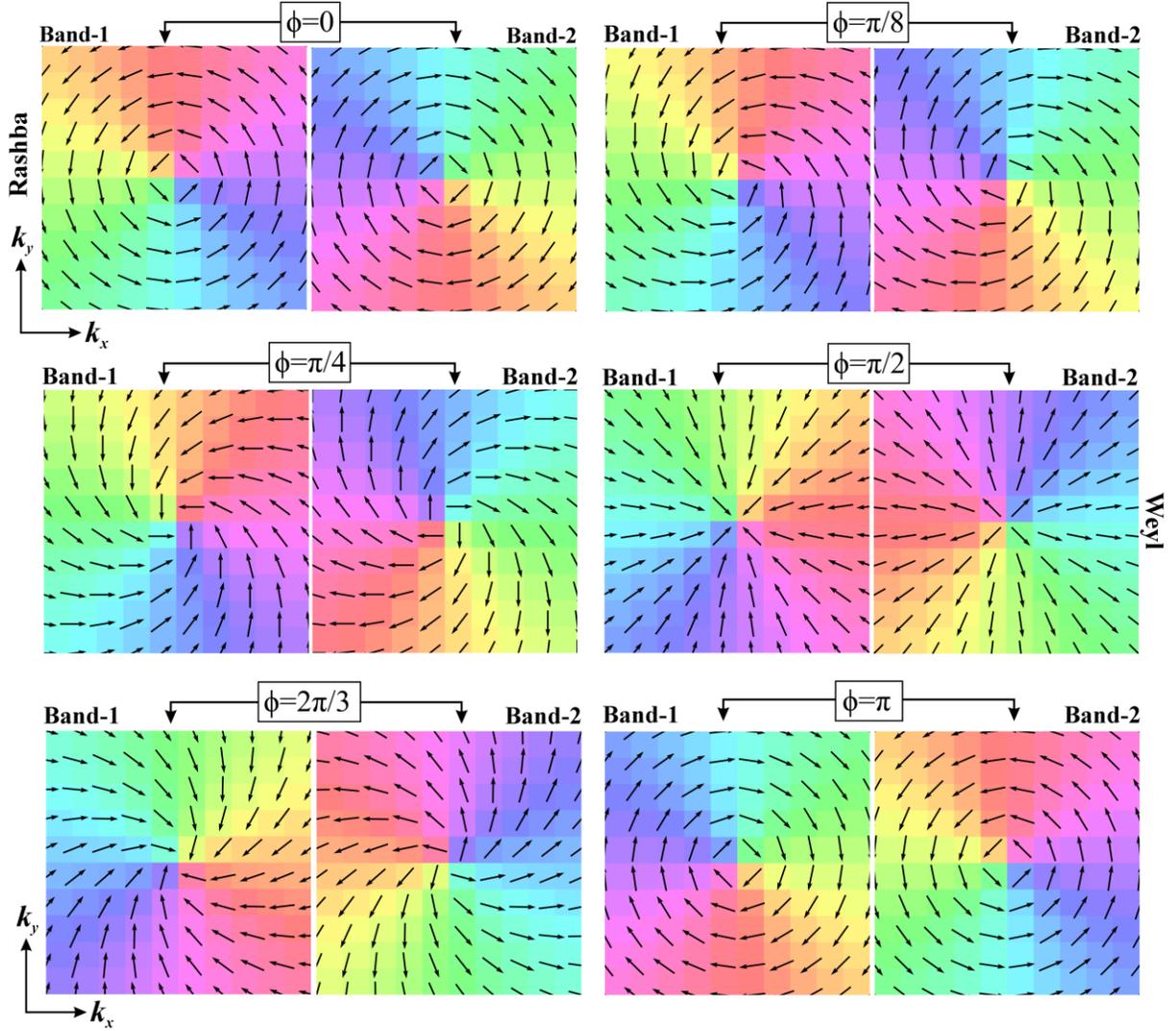

Figure 1: The spin textures obtained from the transformed Hamiltonian $\mathcal{H}_T(k)$ for a different rotation angle $\phi$. For plotting the momentum range of $[k_x, k_y]$ was chosen as [-1,1].

It is noteworthy that similar spiral spin textures have been previously observed when the Rashba and Weyl terms are combined, as reported by Mohanta and Jena in ref. [19].

$$\mathcal{H}_{MJ2} = \alpha(\sigma_x k_y - \sigma_y k_x) + \mathcal{J}(\sigma_x k_x + \sigma_y k_y) \quad \ldots\ldots\ldots (10)$$

For comparison, the spin textures of $\mathcal{H}_{MJ2}$ are shown in Figure 2. Now, comparing Figure 1 and Figure 2, one can conclude that the spin textures obtained from $\mathcal{H}_T(k)$ can be precisely obtained from $\mathcal{H}_{MJ2}$ as well, which makes these two Hamiltonians equivalent.

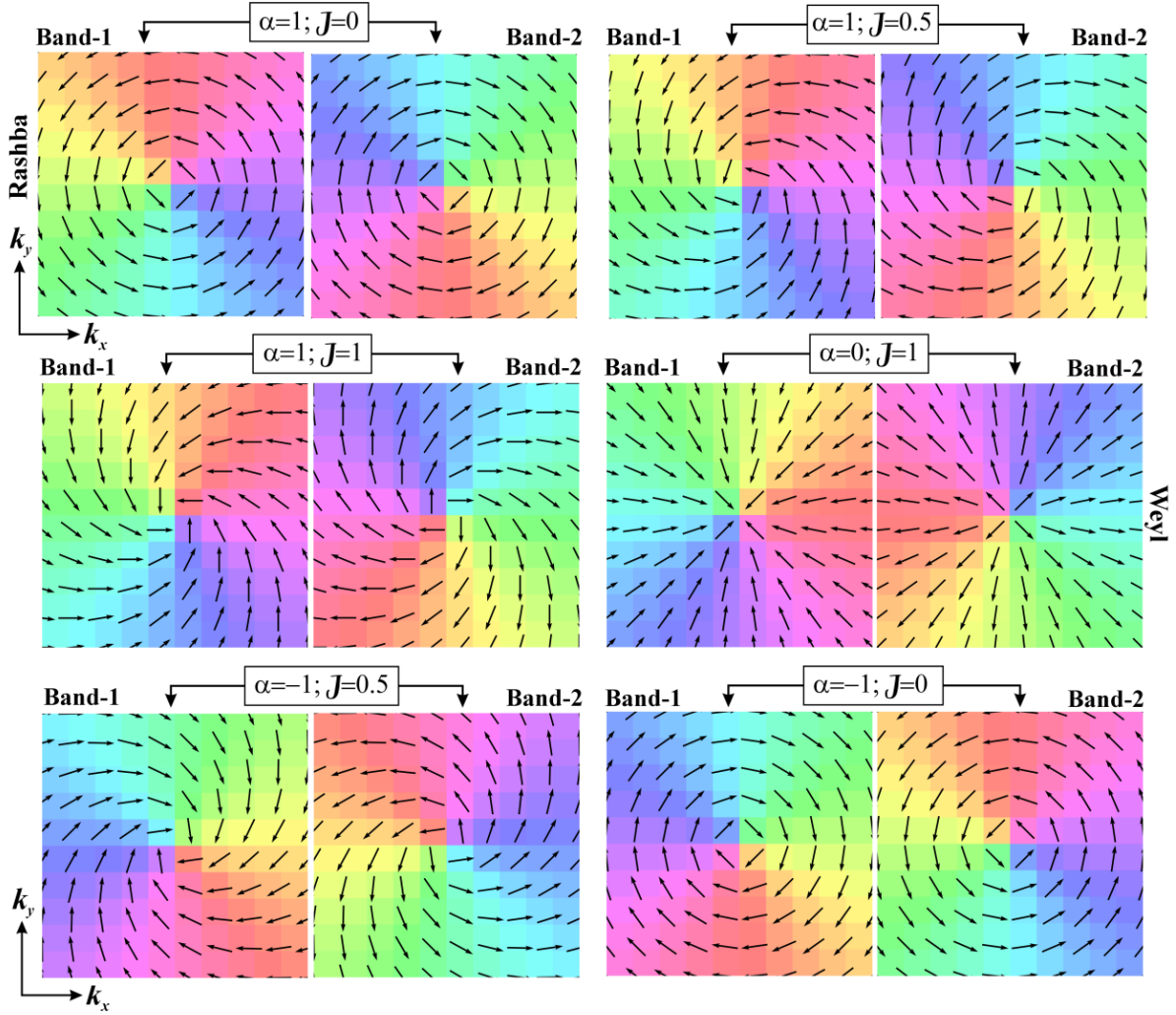

Figure 2: Spin textures obtained from the Hamiltonian $\mathcal{H}_{MJ2}$ for different values of SOC constants. For plotting the momentum range of $[k_x, k_y]$ was chosen as [-1,1].

**2.2 Correspondence between Dresselhaus-1 and Dresselhaus-2 Models:** Using the same unitary operator $(U(\phi) = e^{-i\frac{\phi}{2}\sigma_z})$ and under the same rotation angle $\phi = \frac{\pi}{2}$, the Dresselhaus-1 can be transformed into the Dresselhaus-2 SOC term.

$$\mathcal{H}_{Dresselhaus-1} = (\sigma_x k_x - \sigma_y k_y) \quad \ldots\ldots\ldots\ldots (11)$$

$$\mathcal{H}_{Dresselhaus-2} = (\sigma_x k_y + \sigma_y k_x) \quad \ldots\ldots\ldots\ldots (12)$$

The specific unitary operation that maps one spintronic model onto another is termed as the MKM transformation.

## 3. Application: A case study on Persistent Spin Texture (PST):

Over the past two decades, theoretical investigations have predicted the emergence of a persistent spin texture (PST) under highly restrictive conditions, most notably when the Rashba and Dresselhaus spin-orbit coupling strengths satisfy $\alpha = \pm\beta$ relation. This seminal concept was introduced by Schliemann et al. [20] and has since served as the primary theoretical framework for describing PST. More recently, Mohanta and Jena [19] proposed a related analytical model in which PST can arise when $\alpha = \pm\mathcal{M}$. Collectively, these studies suggest that the realization of PST is limited to narrowly defined parameter regimes, highlighting the stringent conditions traditionally considered necessary for its emergence. These conditions are obtained by considering the following model Hamiltonians:

$$\mathcal{H}_S = \alpha(\sigma_x k_y - \sigma_y k_x) + \beta(\sigma_x k_x - \sigma_y k_y) \quad \ldots\ldots\ldots\ldots\ldots (13)$$

$$\mathcal{H}_{MJ1} = \alpha(\sigma_x k_y - \sigma_y k_x) + \mathcal{M}(\sigma_x k_y + \sigma_y k_x) \quad \ldots\ldots\ldots\ldots (14)$$

Under such extreme conditions, the spin orientation becomes momentum-independent, resulting in an infinite spin lifetime, which is ideal for the design of a nonballistic spin field-effect transistor.

Using the unitary transformation operator defined in the previous section, an alternative formalism can be introduced, as given by;

$$\mathcal{H}_{MKM} = \alpha(\sigma_x k_y - \sigma_y k_x) + \beta(\sigma_x k_x - \sigma_y k_y) + \gamma_1[U(\phi)(\sigma_x k_x + \sigma_y k_y)U(\phi)^\dagger] + \gamma_2[U(\phi)(\sigma_x k_y + \sigma_y k_x)U(\phi)^\dagger] \quad \ldots\ldots\ldots\ldots (15)$$

where $\alpha$, $\beta$ represent the SOC strength of the Rashba and Dresselhaus-1 interactions, whereas $\gamma_1$ and $\gamma_2$ are represents the strength of Weyl and Dresselhaus-2 SOC interactions. Equation (15) provides a unified expression that incorporates the linear contributions from all major spintronic models. The third term maps the Weyl onto the Rashba form (first term), whereas the fourth term converts the Dresselhaus-2 onto the Dresselhaus-1 form (second term) at $\phi = \frac{\pi}{2}$.

Under the condition of $\phi = \frac{\pi}{2}$, $\mathcal{H}_{MKM}$ can be rewritten as:

$$\mathcal{H}_{MKM} = (\alpha - \gamma_1)(\sigma_x k_y - \sigma_y k_x) + (\beta - \gamma_2)(\sigma_x k_x - \sigma_y k_y) \quad \ldots\ldots\ldots (16)$$

Now, for realizing a persistent spin texture (PST), the essential condition obtained for equation (16) is $(\alpha - \gamma_1) = \pm(\beta - \gamma_2)$. The representative parameter sets $[\alpha, \gamma_1; \beta, \gamma_2]$ that satisfy this relation such as [1,0.5; 0.8,0.3], [1.0,0.5; 0.6,0.1] and so on. Notably, these examples demonstrate that the stringent condition of $\alpha = \pm\beta$ is not required for observation of PST as reported earlier, but depends on additional factors as well. Thus, the Hamiltonian $\mathcal{H}_{MKM}$ offers greater flexibility than conventional models, relaxing the otherwise stringent constraints on SOC strengths. It further reveals that PST can originate from broader physical mechanisms, extending beyond the usual constraints imposed by SOC relations. The resulting spin textures obtained from $\mathcal{H}_{MKM}$ for various SOC strengths are shown in Figure 3. It is worth noting that these conditions can be changed depending on how the individual Hamiltonians and the unitary transformation of Hamiltonian $\mathcal{H}_T(k)$ are defined (see SM for more details). The mathematical models for PST are listed in Table 1.

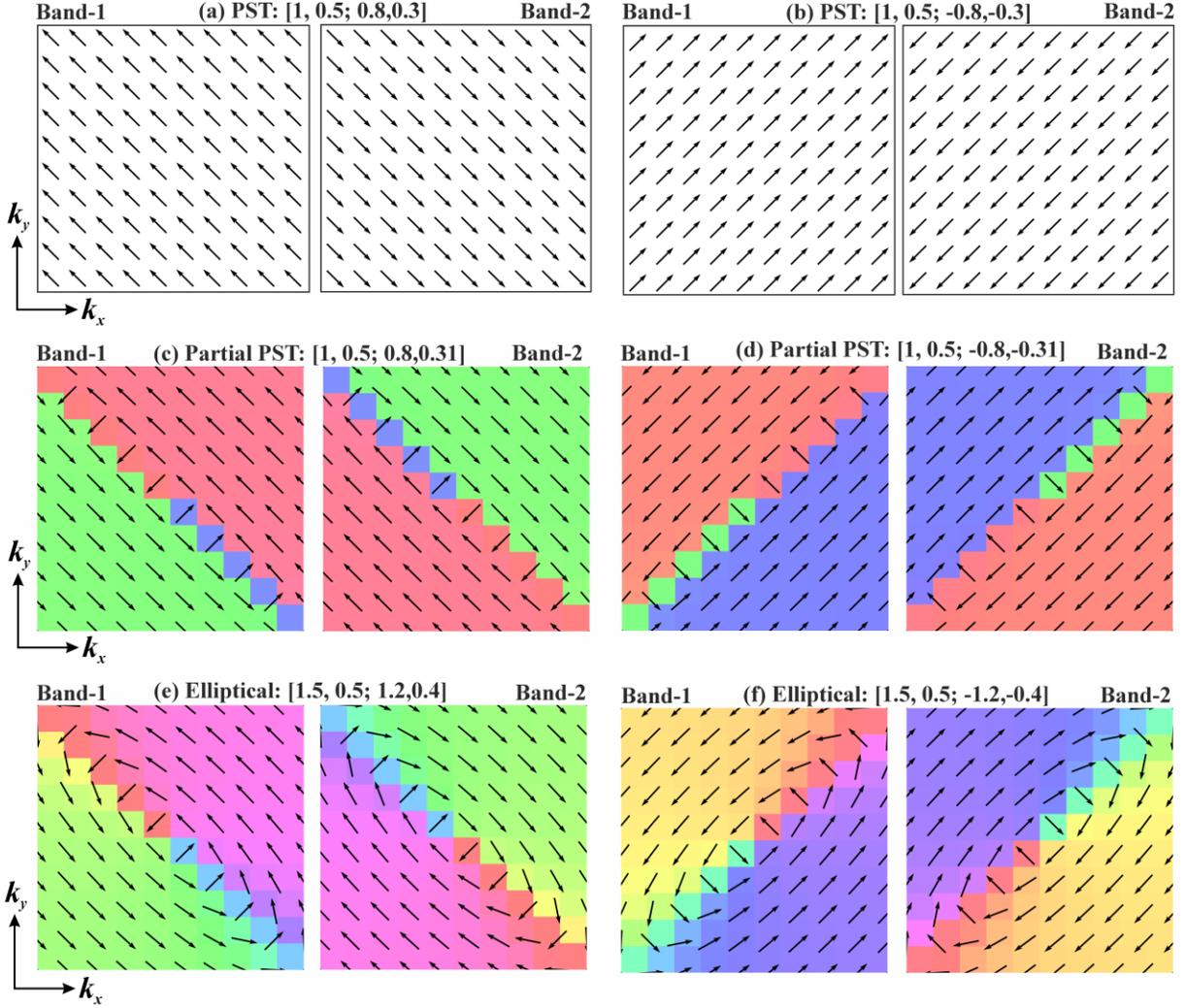

Figure 3: Spin textures obtained from the Hamiltonian $\mathcal{H}_{MKM}$ demonstrating its capability to reproduce the spin textures reported in earlier work: (a-b) PST under the condition of $|\alpha - \gamma_1| = |\beta - \gamma_2|$, (c-d) Partial PST when $|\beta - \gamma_2| \to |\alpha - \gamma_1|$, (d) Elliptical spin pattern when $|\beta - \gamma_2| \ll |\alpha - \gamma_1|$. Similar conditions have been pictorially shown in the references [19,20]. For plotting, the momentum range of $[k_x, k_y]$ was chosen as [-1,1].

**Table 1: Mathematical formulations for the manifestation of persistent spin texture**

| SOC interactions | Conditions for PST | Reference |
|---|---|---|
| $C * [\sigma_x k_y, \sigma_x k_x, \sigma_y k_x, \sigma_y k_y, \sigma_z k_x, \sigma_z k_y]$ | For any value of SOC strength $C$ | [19] |
| $\alpha(\sigma_x k_y - \sigma_y k_x) + \beta(\sigma_x k_x - \sigma_y k_y)$ | $\alpha = \pm \beta$ | [20] |
| $\alpha(\sigma_x k_y - \sigma_y k_x) + \mathcal{M}(\sigma_x k_y + \sigma_y k_x)$ | $\alpha = \pm \mathcal{M}$ | [19] |
| $\alpha(\sigma_x k_y - \sigma_y k_x) + \beta(\sigma_x k_x - \sigma_y k_y) + \gamma_1 [U\left(\frac{\pi}{2}\right)(\sigma_x k_x + \sigma_y k_y)U\left(\frac{\pi}{2}\right)^\dagger]$ | $(\alpha - \gamma_1) = \pm \beta$ | This work (Find more conditions in SM) |
| $\alpha(\sigma_x k_y - \sigma_y k_x) + \beta(\sigma_x k_x - \sigma_y k_y) + \gamma_2 [U\left(\frac{\pi}{2}\right)(\sigma_x k_y + \sigma_y k_x)U\left(\frac{\pi}{2}\right)^\dagger]$ | $\alpha = \pm(\beta - \gamma_2)$ | |
| $\alpha(\sigma_x k_y - \sigma_y k_x) + \beta(\sigma_x k_x - \sigma_y k_y)$ $+ \gamma_1 [U\left(\frac{\pi}{2}\right)(\sigma_x k_x + \sigma_y k_y)U\left(\frac{\pi}{2}\right)^\dagger]$ $+ \gamma_2 [U\left(\frac{\pi}{2}\right)(\sigma_x k_y + \sigma_y k_x)U\left(\frac{\pi}{2}\right)^\dagger]$ | $(\alpha - \gamma_1) = \pm(\beta - \gamma_2)$ | |

Unitary operator: $U(\phi) = e^{-i\frac{\phi}{2}\sigma_z}$

$\alpha$: Rashba SOC strength

$\beta$: Dresselhaus-1 SOC strength

$\gamma_1$: Weyl SOC strength

$\mathcal{M}$ or $\gamma_2$: Dresselhaus-2 SOC strength

**Conclusion:** In conclusion, this study establishes a quantum correspondence between the Rashba and Weyl Hamiltonians, despite their conventional treatment as models for distinct materials and contrasting spin textures. The analytical derivation and graphical analysis show that a unitary rotation in spin space bridges these two models. Further, the same unitary can be applied to map the Dresselhaus-1 onto the Dresselhaus-2 model. This work uncovered an intrinsic rotational symmetry underlying different spin-orbit coupling models. Using the defined unitary transformation, this work proposes a unique analytical model $\mathcal{H}_{MKM}$, consisting of linear terms of all the foundational spintronic models. Supported by analytical derivations and graphical interpretations, unlike existing models, the proposed theory demonstrates that the conventional requirement of equal Rashba ($\alpha$) and Dresselhaus ($\beta/\mathcal{M}$) SOC strength is not the necessary condition for realizing PST.

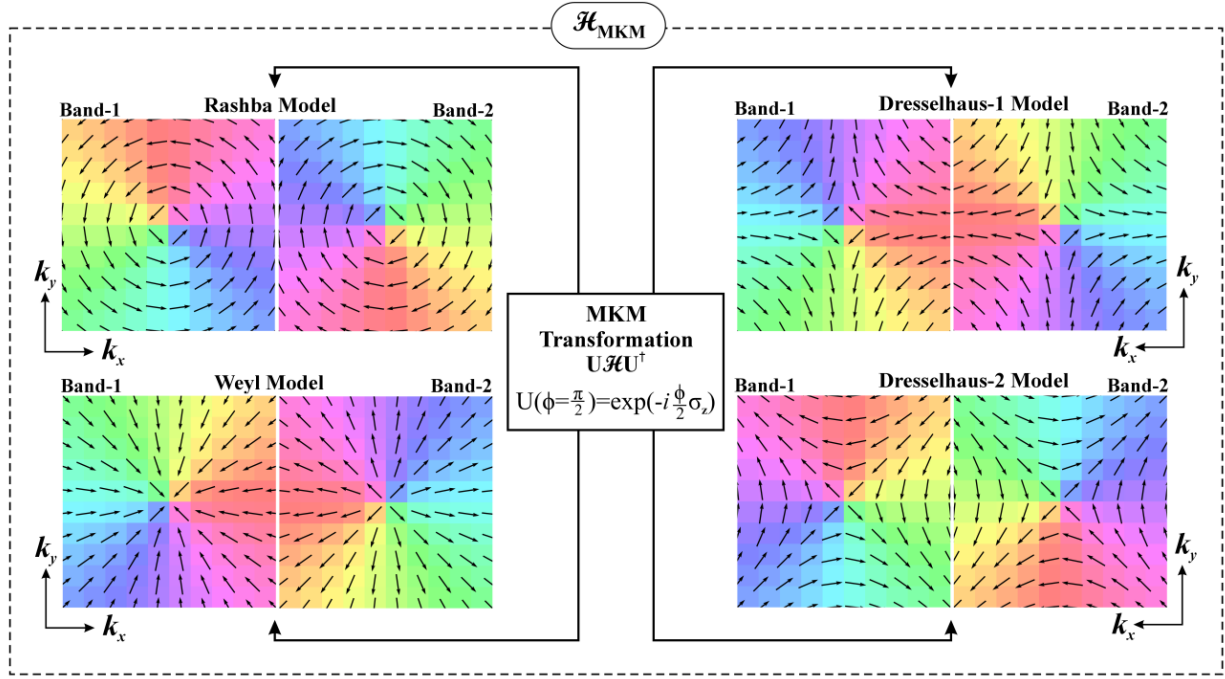

Figure 4: Pictorial representation of the conclusion.

**Acknowledgements:** M. K. Mohanta acknowledges financial support from Anusandhan National Research Foundation (ANRF) under the Ramanujan Faculty Fellowship (Grant No. RJF/2025/000601). The author gratefully acknowledges the research infrastructure and facilities provided by the Indian Institute of Technology-Bhubaneswar.

**Data availability statement:** All the data supporting this work can be found within the paper.